\documentstyle[12pt]{article}
\begin{document}
\newcommand{\be}{\begin{equation}}
\newcommand{\ee}{\end{equation}}
\begin{titlepage}
\title{The redshift periodicity of galaxies \\ as a probe of
the correctness of general relativity\footnote{This paper is based on a 
essay selected for an
Honorable Mention by the Gravity Research Foundation, 1996.}}
\author{Valerio Faraoni 
\\ \\{\small \it Department of Physics and Astronomy, University 
of Victoria} \\
{\small \it P.O. Box 3055, Victoria, B.C. Canada V8W 3P6}\\
{\small \it Tel: [+] 604 721 8656, Fax: [+] 604 721 7715}\\
{\small \it e--mail: faraoni@uvphys.phys.uvic.ca}
}
\date{} 
\maketitle   \thispagestyle{empty}  \vspace*{1truecm}
\begin{abstract}
Recent theoretical work determines the correct coupling constant of 
a scalar field to the 
Ricci curvature of spacetime in general relativity. The periodicity in the 
redshift distribution of galaxies observed by Broadhurst {\em et al.}, 
if genuine, determines the coupling
constant in the proposed scalar field models. As a result, these 
observations contain important information on the
problem whether general relativity is the correct theory of gravity in the 
region of the universe at redshifts $z<0.5$. 
\end{abstract}
\vspace*{1truecm} 
\begin{center} {\small To appear in {\em General Relativity and Gravitation}}
\end{center}

\end{titlepage}   \clearpage

The deep pencil--beam survey of the redshift distribution of galaxies at the
Galactic poles by Broadhurst, Ellis, Koo and Szalay
\cite{BEKS}--\cite{Szalayetal} has detected 
an apparent periodicity in the galaxy distribution with a period 128~$h^{-1}$
Mpc (here $H_0=100$~$ h $~km~s$^{-1}$Mpc$^{-1}$ is the present value of 
the Hubble parameter). The galaxy sample and the statistics used in \cite{BEKS}
has been re--analyzed \cite{KaiserPeacock}--\cite{Dekeletal} and
the result in \cite{BEKS} has been criticized.  Additional data collected 
in the same directions confirmed the original results, supporting the
idea that the peaks in the redshift distribution of galaxies are true features
and not an artifact of inappropriate statistics \cite{Szalayetal}.
Most of the peaks in the galaxy redshift distribution reported in 
\cite{BEKS} have been recognized and confirmed in
\cite{Willmeretal}. The redshift survey at the Keck Telescope discussed in 
\cite{CHPB} also found evidence for strong clustering of galaxies in redshift
space, although the redshift periodicity was not confirmed. 
If the 
periodicity in the redshifts of galaxies is real, one expects a similar 
phenomenon to occur also for the redshifts of Ly~$\alpha$
absorption lines and for Mg~II quasar absorption systems; evidence for
periodicity in the redshift distribution of the Lyman--$\alpha$ forest was
found in \cite{ChuZhu} but not in later observations 
\cite{Bartlettetal,Tytleretal}, and it was argued that it is not genuine
\cite{Scott}. The controversy is reminiscent of the 
old debate about the periodicity of the redshifts of quasars 
\cite{Karlsson}--\cite{Holbaetal}, or of 
similar claims for galaxies \cite{Tifft,GuthrieNapier}. 
Although many authors contend that the result in \cite{BEKS} is probably an 
artifact of incomplete sampling, it
is generally recognized that further observations (expecially 
in directions different 
from the Galactic poles) are needed to definitely resolve the issue, and new
surveys are planned or in progress \cite{newsurveys,CHPB}. It is
established that galaxies tend to cluster in sharp walls, leaving vast regions
devoid of galaxies \cite{GellerHuchra}--\cite{BellangerLapparent}; however, a
single distinguished scale as found in \cite{BEKS} has not been confirmed.
On the other hand, 
there is no doubt that, if confirmed, the claimed redshift periodicity would 
be of the outmost importance, and therefore it has drawn the attention of 
theoreticians \cite{Morikawa}--\cite{SSQ}. 
In order to explain the periodicity found in \cite{BEKS}--\cite{Szalayetal}, 
models were proposed in which clustering of galaxies in
foam--like structures occurs at the predicted redshifts
\cite{vandeWeygaert,IkeuchiTurner}. A 
difficulty of these models is the 
implication that galaxies be approximately distributed on shells, 
of which we happen to be at the center, in conflict with the cosmological
principle. 
 A model in which the cosmological principle is explicitly violated
was proposed recently \cite{Budinichetal}, but it 
is non--viable. In fact, it
assumes that the universe is represented by a closed Friedmann model
dominated by a massless, minimally coupled, non self--interacting scalar field 
$\psi$. The energy density $\rho$ and pressure $P$ of this field are given by 
\be  
\rho=P=\frac{1}{2} \,\nabla_{\alpha}\psi \nabla^{\alpha} \psi \; .
\ee
The stiff equation of state $P=\rho$ certainly does not describe our present,
matter--dominated universe, in which the peculiar velocities of
galaxies give a negligible contribution to the pressure of the cosmological 
fluid ($P=0$). A viable version of the model in
\cite{Budinichetal} would necessarily have to include a mass and/or 
a non--minimal coupling for the scalar field. Such a model would
be similar to the ``oscillating universe'' models introduced in 
\cite{Morikawa,HillSteinhardtTurner}, in which 
the redshift periodicity of galaxies
is an optical illusion rather than a real mass distribution. 
These other models are based on an oscillating scalar field
non--minimally coupled to the Ricci curvature of spacetime.  
We show that if the redshift periodicity of the 
galaxy distribution is real, then the observations contain important
information on the problem whether general relativity is the correct theory 
of gravity. This result arises from the 
observational determination of the coupling constant between the scalar 
field and the curvature \cite{Morikawa,HillSteinhardtTurner}, 
and from recent theoretical 
work on the correct coupling in metric theories of gravity
\cite{SonegoFaraoni,GribPoberii}.

In order to explain the observed distribution of galaxies, Morikawa
\cite{Morikawa} postulated a universe in which the dark matter is 
dominated by a massive, non--minimally coupled, scalar field. His model
was shown be non--viable by Hill, Steinhardt and Turner
\cite{HillSteinhardtTurner}. They proposed 
three alternative models involving oscillating physics: a) an oscillating
``constant'' of gravitation; b) oscillating atomic lines due to time 
variations in the fine structure constant, or in the electron mass; c)
oscillating galactic luminosities. Model b) was ruled out by 
Sudarsky \cite{Sudarsky} on the basis of the Braginsky--Panov experiment. 
Model a) was shown to be compatible with biological records on the Earth 
\cite{SisternaVucetich}, and has recently received renewed interest 
\cite{SSQ}.

The common feature underlying the models under consideration is a spatially 
coherent, oscillating scalar field non--minimally coupled to the Ricci 
curvature $R$ of spacetime. The Lagrangian density of the theory is given
by\footnote{We use the metric signature --~+~+~+. The Riemann tensor is 
given in
terms of the Christoffel symbols $\Gamma_{\mu\nu}^{\rho}$ by 
${R_{\mu\nu\rho}}^{\sigma}=\Gamma_{\mu\rho ,\nu}^{\sigma}-
\Gamma_{\nu\rho ,\mu}^{\sigma}+$~... and the Ricci tensor is
$R_{\mu\rho}= {R_{\mu\nu\rho}}^{\nu}$. $\nabla_{\mu}$ is the 
covariant derivative operator.}
\be
{\cal L}=\frac{1}{16\pi G_0}\, R -\frac{1}{2} 
\nabla_{\mu}\phi \nabla^{\mu}\phi
-\frac{\xi}{2} \, R\,\phi^2-V( \phi)+{\cal L}_m \; ,
\ee
where $G_0$ is the present value of the gravitational ``constant'', 
$\xi$ is the
coupling constant of the scalar field, $V( \phi)$ is its self--interaction 
potential, 
and ${\cal L}_m$ is the Lagrangian density of matter other than the scalar
field. 
The non--minimal ($\xi \neq 0$) coupling seems
to be a common feature of models unifying gravity with the other interactions
and is an essential ingredient of many works on inflation. Two choices of
$\xi$ are most often found in the
literature: $\xi=0$ (``minimal coupling'') and $\xi=1/6$ (``conformal
coupling''). It must 
be emphasized that the correct value of the coupling constant is unknown, 
which can only be regarded as a major piece of missing information regarding 
the physics of the scalar field. Recently, it was
found \cite{SonegoFaraoni} that if $\phi$ is a 
non--gravitational field in a curved spacetime, the Einstein equivalence 
principle (see \cite{Will} for a precise formulation and a discussion) 
forces the value of $\xi$ to be $1/6$. This unintuitive result follows from 
the study of the
Green's function representation of the solutions of the Klein--Gordon equation
\be       \label{KG}
\Box \phi -\xi R \phi-\frac{dV}{d\phi}=0 \; ,
\ee
and from the requirement that, in the neighbourhood of any point of spacetime, 
the propagation 
of scalar waves resembles more and more closely the propagation in flat 
spacetime as the size of the neighbourhood becomes smaller and 
smaller. This is the Einstein
equivalence principle \cite{Will} applied to the propagation of scalar waves.
The quoted result (later confirmed in \cite{GribPoberii}) 
holds for any metric and its derivation is completely
independent of conformal transformations, or of the conformal structure 
of spacetime, or of the field equations of the theory of gravity
\cite{SonegoFaraoni}.
If the nature of the field $\phi$ is gravitational (e.g. the scalar 
field of Brans--Dicke
theory), the statement that its physics resembles {\em locally} the physics 
in flat spacetime goes beyond the Einstein equivalence principle, which is
a statement regarding only {\em non--gravitational} physics \cite{Will} (see
\cite{FaraoniPRD} for a discussion). In
addition, the equation satisfied by a gravitational scalar field may be
different from eq.~(\ref{KG}).

Metric theories of gravity \cite{Will} (including general relativity) 
satisfy the Einstein equivalence principle. Therefore, if the 
correct theory of
gravity describing the region of the universe at redshifts $z<0.5$ 
observed by 
the survey in \cite{BEKS}--\cite{Szalayetal} is general 
relativity, or a metric theory in which the scalar field advocated 
by Hill, Steinhardt and Turner \cite{HillSteinhardtTurner} 
is of non--gravitational origin, then the Einstein
equivalence principle holds, and the coupling constant $\xi=1/6$, as explained
in \cite{SonegoFaraoni}.

Another prescription for $\xi$ is found in a wide class of theories including
Kaluza--Klein, higher derivative, supergravity and superstring--inspired
theories. These theories are originally formulated in the 
``Jordan frame'' (see
\cite{MagnanoSokolowski} for definitions and the terminology) and are
subsequently reformulated in the ``Einstein frame'' by means of a conformal
transformation. The necessity and uniqueness of the conformal transformation
were established for Brans--Dicke and Kaluza--Klein theories 
\cite{Cho,conftrans} and later proved for a wider class of theories
\cite{MagnanoSokolowski}. As a consequence of the conformal transformation to
the Einstein frame, the conformally transformed scalar field is always
minimally coupled to the Ricci curvature of spacetime\footnote{To prevent a
misunderstanding let us mention at this point that Einstein's theory with a
minimally coupled scalar field and Einstein's theory with a conformally 
coupled scalar field are also conformally equivalent to each other (cf. e.g.
\cite{Deser}).} $\xi=0$.

If the scalar field $\phi$ has a quantum nature, other prescriptions apply
according to the nature of $\phi$. If $\phi$ is a Goldstone boson, $\xi=0$
\cite{VoloshinDolgov}; if $\phi$ is the composite boson of the
Nambu--Jona--Lasinio model, the value $\xi=1/6$ was found in the large $N$
approximation \cite{HillSalopek}; if $V( \phi)=\eta \phi^3 /6$, then $\xi=0$
\cite{Hosotani}, while $\xi \leq 0$ or $\xi \geq 1/6$ for Higgs scalar fields
in the standard model \cite{Hosotani}. The value of $\xi$ in a quantum theory
is also affected by renormalization \cite{renormaliz}. In this paper, 
we limit
ourselves to a {\em classical} scalar field $\phi$. In fact, 
the universe at the epoch of galaxy formation was classical, and it 
is believed that also the scalar 
field driving inflation (with which the field $\phi$ has been identified
\cite{Morikawa}) can also be described classically, apart from
the small quantum perturbations generating seeds for the formation of
structures.

It is significant that all the viable models proposed in order to describe the 
observed
distribution of galaxies are based on a classical, oscillating, scalar field 
non--minimally coupled to gravity, and that the value of the constant 
$\xi$ is a
parameter of the models, which can be determined by the observations. 
Typically, the period of the oscillations in redshift of the galaxy 
distribution (128~$h^{-1}$ Mpc) determines the mass of the scalar 
field, while 
the amplitude of the oscillations determines the coupling constant $ \xi$ 
\cite{Morikawa,HillSteinhardtTurner}. It is also possible 
to employ a massless
$\phi$; a variation of the model with oscillating gravitational 
``constant'' predicts $\xi \approx 6.4 \cdot 10^3 $ 
\cite{HillSteinhardtTurner}. An unambiguous fitting of 
the observational data with a value of $\xi$ significantly different from 
$1/6\simeq 0.166 $ implies that general relativity is not the correct 
theory of gravity 
in the observed region of the universe $z<0.5$, because the Einstein
equivalence principle is violated. This would be a remarkable result, since
Solar System experiments seem to indicate that general relativity is the most
plausible theory of gravity \cite{Will}. There remains two possibilities:\\
i) $\phi$ is a gravitational field in a metric theory of gravity other than
general relativity: in this case the Einstein equivalence principle and 
the prescription $\xi=1/6$ do not apply to $\phi$; \\
ii) gravity is described by 
a non--metric theory of gravitation, in which the Einstein equivalence 
principle (and consequently the prescription $\xi=1/6$) does not necessarily 
hold.

The latter possibility is regarded as highly
unlikely, since there are good reasons to believe that the correct 
theory of gravity is a
metric theory \cite{Will}. If case i) is appropriate, the observational 
determination of the value of $\xi$ is not sufficient to single 
out the correct
theory of gravity, but can nevertheless provide a important hint. If 
one of the
models in \cite{HillSteinhardtTurner} is correct, then $\xi
\neq 0$ and the correct (low--energy) theory of gravity is {\em not} one
formulated in the Einstein frame: this would exclude Brans--Dicke,
Kaluza--Klein, supergravity and superstring theories. In addition, a 
reliable determination of $\xi$, which is otherwise unknown, is regarded as 
a remarkable progress in the physics of the scalar field.

The spatial coherence of the scalar field $\phi$ has been related to
inflation \cite{HillSteinhardtTurner}, and the scalar field $\phi$ has been
explicitly identified with the inflaton \cite{Morikawa}. 
If $\phi$ is to be 
identified with the field driving
inflation, the value of the coupling constant has a
profound effect on the success of the inflationary scenario under
consideration \cite{FaraoniPRD}. Since the 
proposed inflationary 
scenarios are currently under test using observations of the 
cosmic microwave background, a reliable measurement of the coupling constant
$\xi$ would give further hints on the correctness of some of the proposed
scenarios for inflation. However, it is to be stressed that the region of the
universe under study (redshifts $z<0.5$) does not correspond to the very early
universe, which is observationally inaccessible, but it is well within the
range of present astronomy.

Ultimately, the problems whether one of the oscillating universe models is 
correct and whether it is possible to measure the coupling constant 
$\xi$ are meaningful only if
the redshift periodicity discovered in \cite{BEKS}--\cite{Szalayetal} 
is genuine. Since general relativity has
been succesfully tested in the Solar System \cite{Will}, the belief that it is
valid at least in the region of the universe at redshifts $z<0.5$ has a firm 
justification, and one could go as far as considering this fact, in conjunction
with the suggested 
values of $\xi$ ($\xi=10$ \cite{Morikawa}, $\xi=6.4 
\cdot 10^3$ \cite{HillSteinhardtTurner}, $\xi=6.267$ or other values 
\cite{SSQ}), 
as an argument against the authenticity of the redshift periodicity reported in
\cite{BEKS}--\cite{Szalayetal}. While we do
not support this extreme point of view, we believe that the possibility 
of measuring the coupling constant $\xi$ and its implications for the
determination of the correct theory of gravity provide new and compelling 
motivation for resolving the issue of the redshift periodicity of galaxies. To
understand the large--scale clustering of galaxies, new pencil beams have been
analyzed and new surveys are being carried on
\cite{newsurveys,BellangerLapparent}.

\section*{Acknowledgments}
The author acknowledges F.I. Cooperstock, C.J. Pritchet and R. De Propris for
stimulating discussions, and a referee for suggestions leading to improvements
in the manuscript. This work was supported by the NATO Advanced 
Fellowships Programme through the National Research Council of Italy (CNR).

\clearpage          
{\small 
\begin{thebibliography}{99}

\bibitem{BEKS} T.J. Broadhurst, R.S. Ellis, D.C. Koo and A.S. Szalay, 
Nature 343 (1990) 726.

\bibitem{Kooetal} D.C. Koo {\em et al.}, in ASP Conf. Ser. 51, Observational
Cosmology, ed. by G. Chincarini {\em et al.} (ASP, S. Francisco), p.~112.


\bibitem{Szalayetal} A.S. Szalay {\em et al.}, Proc.. Nat. Acad. Sci. 90 (1993)
4853.

\bibitem{KaiserPeacock} N. Kaiser and J.A. Peacock, Astron. J. 379 (1991) 
482.

\bibitem{ParkGott} C. Park and J.R. Gott, Mon. Not. R. Astr. Soc. 249 (1991)
288. 

\bibitem{Dekeletal} A. Dekel, G.R. Blumenthal, J.R. Primack and 
D. Stanhill, Mon. Not. R. Astr. Soc. 257 (1992) 715.

\bibitem{Willmeretal} C.N.A. Willmer {\em et al.}, Astrophys. J. 437 (1994)
560.

\bibitem{CHPB} J.C. Cohen, D.W. Hogg, M.A. Pahre and R. Blandford, 
Astrophys. J. Lett. 462 (1996) L9.

\bibitem{ChuZhu} Y. Chu and X. Zhu, Astron. Astrophys. 222 (1989) 1.

\bibitem{Bartlettetal} J.G. Bartlett, R. Esmailzadeh and L.J. Hall, 
UC Berkeley preprint (1990).

\bibitem{Tytleretal} D. Tytler, J. Sandoval and X.M. Fan, Astrophys. J. 405 
(1993) 57.

\bibitem{Scott} D. Scott, Astron. Astrophys. 242 (1991) 1.

\bibitem{Karlsson} K.G. Karlsson, Astr. Astrophys. 13 (1971) 333. 

\bibitem{Depaquitetal} S. Depaquit, J.C. Pecker and J.P. Vigier, 
Astr. Nachr. 306 (1985) 7.

\bibitem{Arpetal} H. Arp, H.G. Bi, Y. Chu and X. Zhu, Astr. Astrophys.
239 (1990) 33.

\bibitem{Holbaetal} A. Holba {\em et al.}, Astrophys. Sp. Sci. 222 (1994) 65.

\bibitem{Tifft} W.G. Tifft, Astrophys. J. 206 (1974) 38.

\bibitem{GuthrieNapier} B.N.G. Guthrie and W.M. Napier, 
Mon. Not. R. Astr. Soc. 253 (1991) 533. 

\bibitem{newsurveys} D. Fabricant, E. Hertz and A. Szentgyorgy, Proc.
SPIE, 2198 (1994); M.J. Geller, J. Roy. Astr. Soc. Can. 88 (1994) 283; 
G. Vettolani {\em et al.}, in Proc. 9th IAP Meeting, Cosmic Velocity Fields,
ed. by F.R. Bouchet and M. Lachi\`{e}ze--Rey (Editions Frontieres, Gyf sur
Yvette), p.~523.

\bibitem{GellerHuchra}  M.J. Geller and J.P. Huchra, Science 246 (1989) 897.

\bibitem{daCosta} L.N. da Costa {\em et al.}, Astrophys. J. 424 (1994) L1.

\bibitem{BellangerLapparent} C. Bellanger and V. de Lapparent, Astrophys. J.
455 (1995) L103.

\bibitem{Morikawa} M. Morikawa, Astrophys. J. Lett. 362 (1990) L37; 
Astrophys. J. 369 (1991) 20.

\bibitem{HillSteinhardtTurner} C.T. Hill, P.J. Steinhardt and M.S. Turner, 
Phys. Lett. B 252 (1990) 343.

\bibitem{Sudarsky} D. Sudarsky, Phys. Lett. B 281 (1992) 281.

\bibitem{SisternaVucetich} P.D. Sisterna and H. Vucetich, Phys. 
Rev. Lett. 72 (1994) 454.

\bibitem{Budinichetal} P. Budinich, P. Nurowski, R. Raczka and M. Ramella, 
Astrophys. J. 451 (1995) 10; P. Budinich and R. Raczka, 
Found. Phys. 23 (1993) 225; Europhys. Lett. 23 (1993) 295.

\bibitem{SSQ} M. Salgado, D. Sudarsky and H. Quevedo, Phys. Rev. D 53 (1996)
6771.

\bibitem{vandeWeygaert} R. van de Weygaert, Mon. Not. R. Astr. Soc.
249 (1991) 159.

\bibitem{IkeuchiTurner} S. Ikeuchi and E.L. Turner, Mon. Not. R. 
Astr. Soc. 250 (1991) 519.

\bibitem{SonegoFaraoni} S. Sonego and V. Faraoni, Class. Quant. 
Grav. 10 (1993) 1185.

\bibitem{GribPoberii} A.A. Grib and E.A. Poberii, Helv. Phys. Acta 68 (1995)
380.

\bibitem{Will} C. M. Will, {\em Theory and Experiment in Gravitational
Physics}, revised edition (Cambridge Univ. Press, Cambridge, 1993).

\bibitem{FaraoniPRD} V. Faraoni, Phys. Rev.~D 53 (1996) 6813.

\bibitem{MagnanoSokolowski} G. Magnano and L.M. Sokolowski, Phys.
Rev.~D 50 (1994) 5039.

\bibitem{Cho} Y.M. Cho, Phys. Rev. Lett. 68 (1992) 3133.

\bibitem{conftrans} L. Bombelli {\em et al.}, Nucl. Phys. B 289 (1987) 735; 
L.M. Sokolowski and Z. Golda, Phys. Lett. B 195 (1987) 349; L.M. Sokolowski,
Class. Quant. Grav. 6 (1989) 59; 2045.

\bibitem{Deser} S. Deser, Phys. Lett. B 134 (1984) 419.

\bibitem{VoloshinDolgov} M.B. Voloshin and A.D. Dolgov, Sov. J. Nucl. 
Phys. 35 (1982) 120.

\bibitem{HillSalopek} C.T. Hill and D.S. Salopek, Ann. Phys. (NY)
213 (1992) 21.

\bibitem{Hosotani} Y. Hosotani, Phys. Rev.~D 32 (1985) 1949.

\bibitem{renormaliz} M. Reuter, Phys. Rev.~D 49 (1994) 6379; L. 
Parker and D.J. Toms,  Phys. Rev.~D 32 (1985) 1409.

\end{thebibliography} }
\end{document}